\documentclass[aps,prl,twocolumn,10pt,superscriptaddress,showpacs]{revtex4-1}

\usepackage{graphicx}
\usepackage{hyperref}
\usepackage{colortbl}
\usepackage[dvipsnames]{xcolor}
\usepackage{amsmath,amsfonts}
\usepackage{dcolumn}
\usepackage{ifthen}

\usepackage{xcolor}

\usepackage[normalem]{ulem}
\newcommand{\rem}[1]{\textcolor{red}{\sout{#1}}}
\newcommand{\switch}[1]{%
  \ifthenelse{\equal{#1}{0}}{\renewcommand{\rem}[1]{}}{}}
\switch{0}

\renewcommand\footnotemark{}

\usepackage{fancyhdr}

\fancypagestyle{specialfooter}{%
  \fancyhf{}
  
  \fancyfoot[R]{ Accepted for  publication in PRL: \newline http://journals.aps.org/prl/accepted/de075Y22He31ad4fe8966203671b32084ef612abb \newline \textcopyright\ 2015 American Physical Society \newline}
}

\begin{document}

\title{Emergence and Persistence of Collective Cell Migration on Small Circular Micropatterns}

\author{Felix J. Segerer}
\thanks{Felix J. Segerer and Florian Th\"uroff contributed equally to this work.}
\affiliation{Faculty of Physics and Center for NanoScience Ludwig-Maximilians-Universit\"at M\"unchen, \\ Geschwister-Scholl-Platz 1, D-80539 Munich, Germany}

\author{Florian Th\"uroff}
\thanks{Felix J. Segerer and Florian Th\"uroff contributed equally to this work.}
\affiliation{Arnold-Sommerfeld-Center for Theoretical Physics and Center for NanoScience, Faculty of Physics, \\ Ludwig-Maximilians-Universit\"at M\"unchen, Theresienstrasse 37, D-80333 Munich, Germany
}

\author{Alicia \surname{Piera Alberola}}%
\affiliation{Faculty of Physics and Center for NanoScience Ludwig-Maximilians-Universit\"at M\"unchen, \\ Geschwister-Scholl-Platz 1, D-80539 Munich, Germany}

\author{Erwin Frey}%
\email{frey@lmu.de}
\affiliation{Arnold-Sommerfeld-Center for Theoretical Physics and Center for NanoScience, Faculty of Physics, \\ Ludwig-Maximilians-Universit\"at M\"unchen, Theresienstrasse 37, D-80333 Munich, Germany
}

\author{Joachim O. R\"adler}%
\email{raedler@lmu.de}
\affiliation{Faculty of Physics and Center for NanoScience Ludwig-Maximilians-Universit\"at M\"unchen, \\ Geschwister-Scholl-Platz 1, D-80539 Munich, Germany}


\begin{abstract}
The spontaneous formation of vortices is a hallmark of collective cellular activity. Here, we study the onset and persistence of coherent angular motion (CAMo) as a function of the number of cells $N$ confined in circular micropatterns. We find that the persistence of CAMo increases with $N$ but exhibits a pronounced discontinuity accompanied by a geometric rearrangement of cells to a configuration containing a central cell. Computer simulations based on a generalized Potts model reproduce the emergence of vortex states and show in agreement with experiment that their stability depends on the interplay of spatial arrangement and internal polarization of neighboring cells. Hence, the distinct migrational states in finite size ensembles reveal significant insight into the local interaction rules guiding collective migration.
\end{abstract}

\pacs{87.18.Gh,87.17.Jj,87.18.Fx,87.17.Aa}


\maketitle
\thispagestyle{specialfooter}

The ability of cells to coordinate their motion is essential in various biological contexts, notably morphogenesis \cite{Ewald:2008, Vasilyev:2009, Lecaudey:2006} and tissue repair \cite{Shaw:2009, Poujade:2007}. 
In recent studies, monolayers of Madin-Darby canine kidney (MDCK) cells have been investigated as model systems for collective behavior in living systems. Remarkably large scaled correlations and swirls in cell migration have been observed and characterized using image correlation and traction force microscopy techniques \cite{Poujade:2007, Petitjean:2010, Angelini:2010, Marel:2014}. These emergent patterns and correlations were attributed to cell-cell coupling, and mechano-transduction mediated by the force-generating cytoskeleton. In fact, dynamic self-ordering into streaming patterns and vortex states appears to be rather generic in assemblies of (self-)propelled objects. They are well known in active systems as diverse as driven biopolymers in motility assays~\cite{Schaller:2010, Butt:2010, Sumino:2012}, bacterial colonies~\cite{Dombrowski:2004,Ben-Jacob:2006, Wioland:2013}, and driven granular media~\cite{Narayan:2007, Kudrolli:2008, Weber:2013}. Superficially, these phenomena may be attributed to a tendency of neighboring objects to align their direction of motion, as suggested by flocking models~\cite{Vicsek:1995}. However, upon closer inspection, there are many important qualitative differences between all these systems and to date quantitative theoretical models are largely lacking. 

For cell assemblies, the challenge is that mechanical and biochemical interactions between cells as well as internal organization of cells are complex \cite{Ridley:2003}, and therefore parameter control is limited. Recent progress in understanding collective behavior of cell assemblies has been fueled by micropatterning techniques which enabled well-controlled in-vitro experimental systems. These techniques have been used to study static adherence and intracellular cytoskeleton organization of individual cells in defined geometries~\cite{Thery:2010}. Importantly, geometrical confinement of cells into micropatterned circles has been found to induce persistent rotational motion for systems ranging from two cells~\cite{Huang:2005} to large assemblies~\cite{Doxzen:2013, Deforet:2014}~\footnote{While 
Deforet et al.~\cite{Deforet:2014} observed global changes in rotation direction for wild type MDCK cells on large scale patterns, this was not observed by Doxzen et al.~\cite{Doxzen:2013}. Since substrate stiffness can influence collective behavior significantly~\cite{Angelini:2010,Ng:2012}, this discrepancy might be due to the different substrates used (PDMS/Glass).}. There is general consensus that on a macroscopic scale collective cell migration is to a large degree generic and can be explained by different classes of theoretical models including flocking models~\cite{Vicsek:1995, Vicsek:2012, Basan:2013, Sepulveda:2013}, cellular Potts models \cite{Graner:1992, Glazier:1993, Szabo:2006, Szabo:2010, Kabla:2012}, and phase field models \cite{Camley:2014, Lober:2015}.
However, the mechanisms underlying the emergence of vortex states are still poorly understood and its relationship to single-cell properties remains unclear. In particular, a systematic study of the emergence and stability of small-scale vortex states and the dynamic disorder-order transition leading to the emergence of collective migration as a function of the number of involved cells has not been carried out so far.

\begin{figure}[t]
\includegraphics[width=1\linewidth]{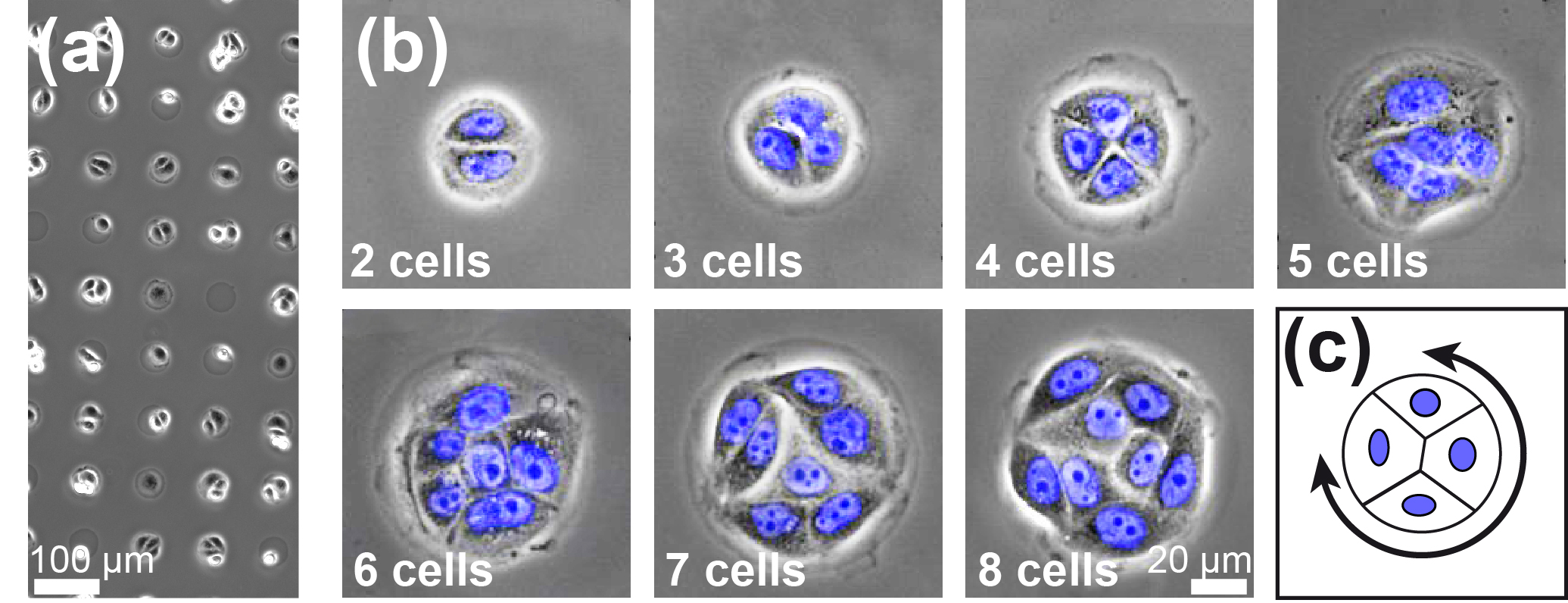}
\caption{\label{Fig.1} (a) Array of MDCK cells seeded on circular micropatterns. (b) Circular patterns occupied by 2-8 cells. Circle size increases in such a way that the average area per cell is constant at approximately $830\; \mathrm{\mu m^2}$. Nuclei are labeled in blue. (c) Schematic of 4 cells rotating within a circular field.
}
\end{figure}

Here, we investigated the emergence of collective rotational motion in small circular micropatterns as a function of the number of cells (Fig.~1). The physical system consists of arrays of circular fields containing 2-8 cells. Cell density is kept constant by increasing field size in line with cell number. We found distinct transitions between states of disordered motion (DisMo) and states of coherent angular motion (CAMo). Furthermore, the survival time of the coherent state tends to increase with increasing cell number, but shows a pronounced drop between 4 and 5 cells, where the geometric cell arrangement changes from a conformation without a cell in the system center to one including a centered cell. Employing a computational model, based on the cellular Potts model (CPM)~\cite{Graner:1992,Glazier:1993}, which we extended to incorporate internal polarization and cell-to-cell mechano-transduction~\cite{Thueroff:2014}, we reproduced and explained these features. Thus, the experimentally observed gradual transition with increasing system size from predominantly erratic motion of small cell groups to directionally persistent migration of larger assemblies is captured by the theory, underlining the role of internal cell polarity in the emergence of collective behavior.

Micropatterns of the extracellular matrix protein fibronectin separated by PEGylated cell-repelling areas were fabricated using a plasma-induced patterning approach. Parts of a culture dish (Ibidi) were covered with a polydimethylsiloxan (PDMS) template of the desired pattern. Exposed parts were treated with $O_2$-plasma in a plasma cleaner (electronic diener) and overlaid for 30~min with 1~$\mathrm{mg/mL}$ PLL(20)-g[3.5]-PEG(2) (SuSoS). Afterwards, the template was removed and the whole surface was briefly exposed to a 50~$\mathrm{\mu g/mL}$ solution of fibronectin (Yo Proteins). MDCK cells were seeded on the structured surface and placed in a temperature-controlled environmental chamber on the microscope stage. Arrays of circles were designed with increasing sizes to accommodate 2-8 cells (Fig.\ref{Fig.1}). To ensure constant cell density of 830~$\mathrm{\mu m^2/cell}$, for each pattern size, only fields containing the appropriate number of cells were selected for analysis. Nuclei were stained using Leibowitz L-15 medium (c-c-pro) containing 15~$\mathrm{ng/mL}$ Hoechst 33342 (Invitrogen). Time lapse movies were recorded at a rate of 6 frames/hour over 50~h using an iMIC automated microscope (TILL Photonics).
Individual nuclei were tracked using in-house image analysis software. 

\begin{figure}
\includegraphics[width=1\linewidth]{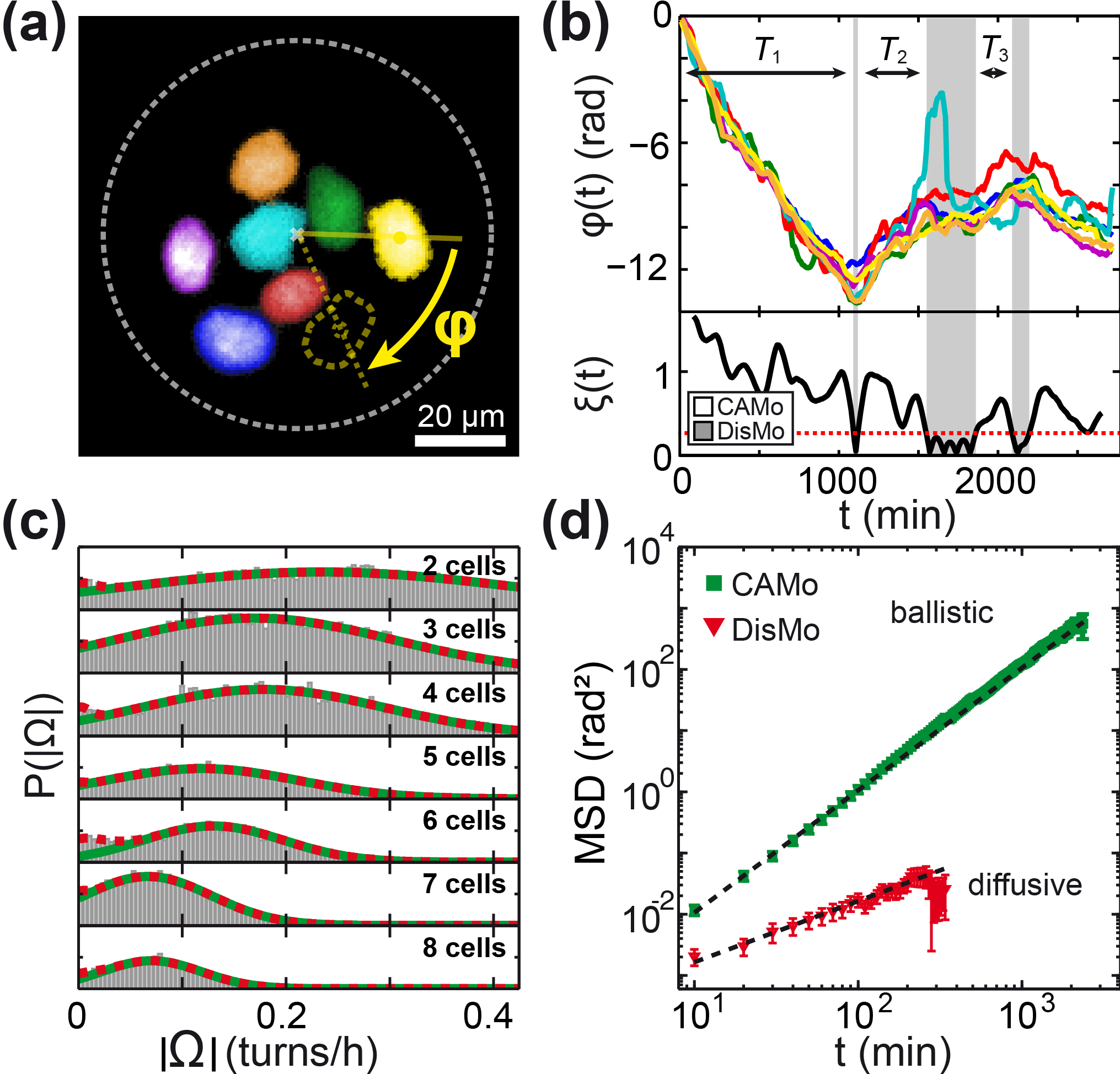}
\caption{\label{Fig.2}(color). (a) False-color fluorescence image of the nuclei of seven cells within a circular micropattern. For each nucleus $i$, the angular position $\varphi_i(t)$ was evaluated with respect to the circle center. (b) Angular positions $\varphi_i(t)$ of each cell (in colors corresponding to the nuclei in (a)) and normalized total angular velocity $\xi(t)$. The classification threshold of $\xi_c=1/4$ is indicated by the red dashed line. Periods of DisMo are highlighted by gray shaded areas. 
(c) Probability distribution of the mean angular velocity $|\Omega_N|$ for systems containing 2 to 8 cells. The distributions are fitted by a single Gaussian (green) and a mixture of two Gaussians (dashed red). The deviation between the two curves reveals a local maximum at $|\Omega_N|=0$.
(d) Log-log plot of the angular MSD of CAMo (green) and DisMo (red) and its error for assemblies consisting of 8 cells. For other cell numbers see Fig.~S4 \cite{Supplement}.
}
\end{figure}


A typical array of circular adhesion sites occupied by MDCK cells is shown in Fig.~\ref{Fig.1}(a). Cells exhibit spontaneous collective rotation within the circular areas (Fig.~\ref{Fig.1}(c)). Periods of CAMo are seen to be interrupted by intervals of DisMo, after which rotation in an arbitrary direction is resumed (for movies see \cite{Supplement}). 
Increasing the system size cell by cell (Fig.~\ref{Fig.1}(b)), we studied collective rotation as a function of cell number. For each cell $i$, the center of the nucleus was tracked and recorded in polar coordinates, and the individual angular positions $\varphi_i(t)$ were calculated (Fig.~\ref{Fig.2}(a)) (for a detailed description see section S2 \cite{Supplement}). Typical time courses of $\varphi_i(t)$ for a system of 7 cells are shown in Fig.~\ref{Fig.2}(b).
To filter out small fluctuations which result, for example, from displacements of the nucleus with respect to the geometric center of the cell, we calculated the system angular velocity $\Omega_N(t)$ as the mean over the individual angular velocities of the $N$-cell system smoothed over a number of frames $n_f$ taken in discrete intervals of $T_f=10\; \mathrm{min}$:
\begin{equation} \label{eq:omega} 
\Omega_N(t) =\frac{1}{N \cdot n_{f}^2 \cdot T_f} \sum_{i=1}^{N} \sum_{\tau=t}^{t+n_f} \lbrack  \varphi_i (\tau) -  \varphi_i (\tau-n_f) \rbrack.
\end{equation}
We chose $n_f=9$ as the best trade-off between smoothing of fluctuations and temporal resolution
\footnote{Data analysis using $n_f=4$ or $n_f=15$ respectively had no significant qualitative influence on the results (data not shown).}.
For all $N$, the probability distribution $P(\Omega_N)$ displays symmetry breaking into clockwise and counterclockwise rotations. Both directionalities are almost equally represented, with a small bias towards clockwise rotation (see Fig.~S2 \cite{Supplement}). Similar chiralities have been reported before \cite{Wan:2011, Vandenberg:2013, Doxzen:2013}.

To distinguish periods of CAMo from periods of DisMo, we analyzed the probability distribution $P(|\Omega_N|)$. It was found to be approximately Gaussian (Fig.~\ref{Fig.2}(c)). The maximum, $\bar{\Omega}_N$, as well as the standard deviation, $\sigma_N$, decreased with increasing cell number, displaying an almost constant coefficient of variation $\sigma_N / \bar{\Omega}_N = 0.74 \pm 0.13$. At $\Omega_N=0$, $P(|\Omega_N|)$ exhibits a weak second maximum, indicating a state of disordered, i.e. non-rotating, motion.
Introducing a normalized variable $\xi_N(t) := \left|  \Omega_N (t)  / \bar{\Omega}_N \right|$, we defined
a common threshold for all $N$ at $\xi_c=1/4$, so that for $\xi_N(t)<\xi_c$ a migration state is classified as DisMo and for $\xi_N(t) \ge \xi_c$ as CAMo respectively (Fig.~S2 \cite{Supplement}).
(As discussed in section S3 \cite{Supplement}, an alternative approach to identify collective motion gave the same results). To verify that these two states are distinct in their migrational behavior we calculated the angular mean squared displacement (MSD) during each state, $
\text{MSD}(t)= \langle [\langle\varphi (t)\rangle_{N} -\langle\varphi (0)\rangle_N]^2 \rangle_{\mathrm{states}}$, where $t=0$ signifies the starting point of an interval. Averages were taken over all $N$ cells within a given system as well as over all observed intervals of CAMo or DisMo, denoted by $\langle \ldots \rangle_N$ and
$\langle \ldots\rangle_{\text{states}}$, respectively.
Consistently, the MSD of CAMo shows a slope 2 in a log-log plot, indicating ballistic angular motion for all cell numbers, while the MSD of DisMo exhibited diffusive behavior (Fig.~\ref{Fig.2}(d)).

\begin{figure}
\includegraphics[width=1\linewidth]{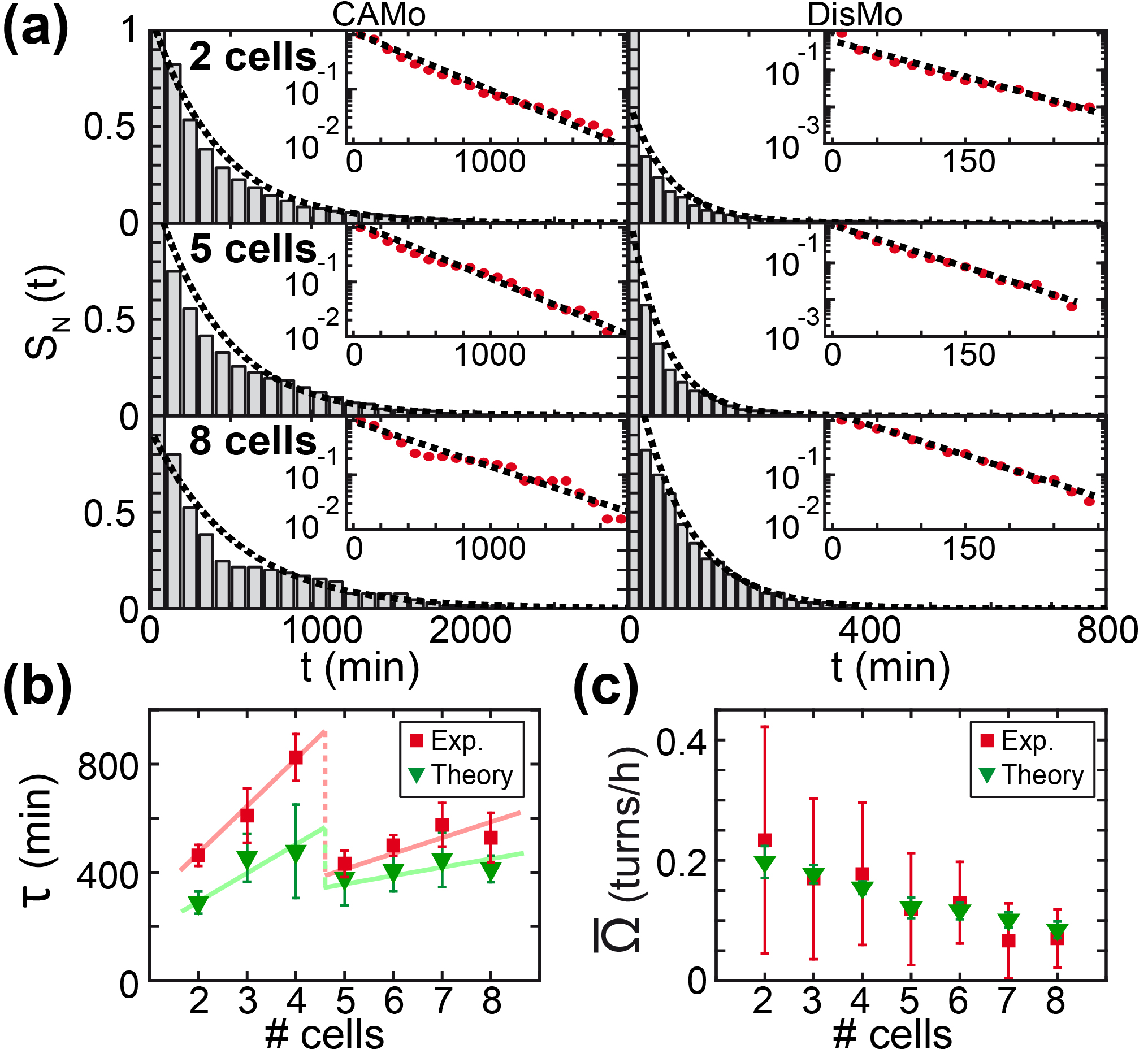}
\caption{\label{Fig.3}(color online). (a) Survival function $S_N(t)=P_N({T>t})$ of CAMo and DisMo states. Insets show corresponding log-lin plots. Exponential fits are indicated by dashed lines (for other cell numbers see Fig.~S5 \cite{Supplement}). (b) Persistence time $\tau$ as a function of cell number, derived from experiment and theory. Error bars indicate confidence bounds of 99\% within the fits. (c) Peak positions $\bar{\Omega}_N$ of the distribution of the angular velocity $P(\Omega_N)$ from experimental data and theory. Error bars indicate the standard deviation.
}\end{figure}

Next we evaluated the lifetimes of the CAMo and DisMo states.
Fig.~\ref{Fig.3}(a) shows the survival probability $S_N(t)=P_N({T>t})$, i.e. the fraction of CAMo/DisMo time periods $T$ exceeding $t$, based on a sample size of over 600 systems (see Table S2 \cite{Supplement}). We found that the survival probabilities of both states decay exponentially, $S_N(t) \propto e^{-t/ \tau}$ suggesting that the stochastic process underlying the emergence and collapse of both states is Poissonian. The persistence time $\tau$ of the coherent state increases with increasing cell number, but exhibits a pronounced discontinuity between systems containing 4 and 5 cells (Fig.~\ref{Fig.3}(b)). 
\begin{figure*}
\includegraphics[width=1\linewidth]{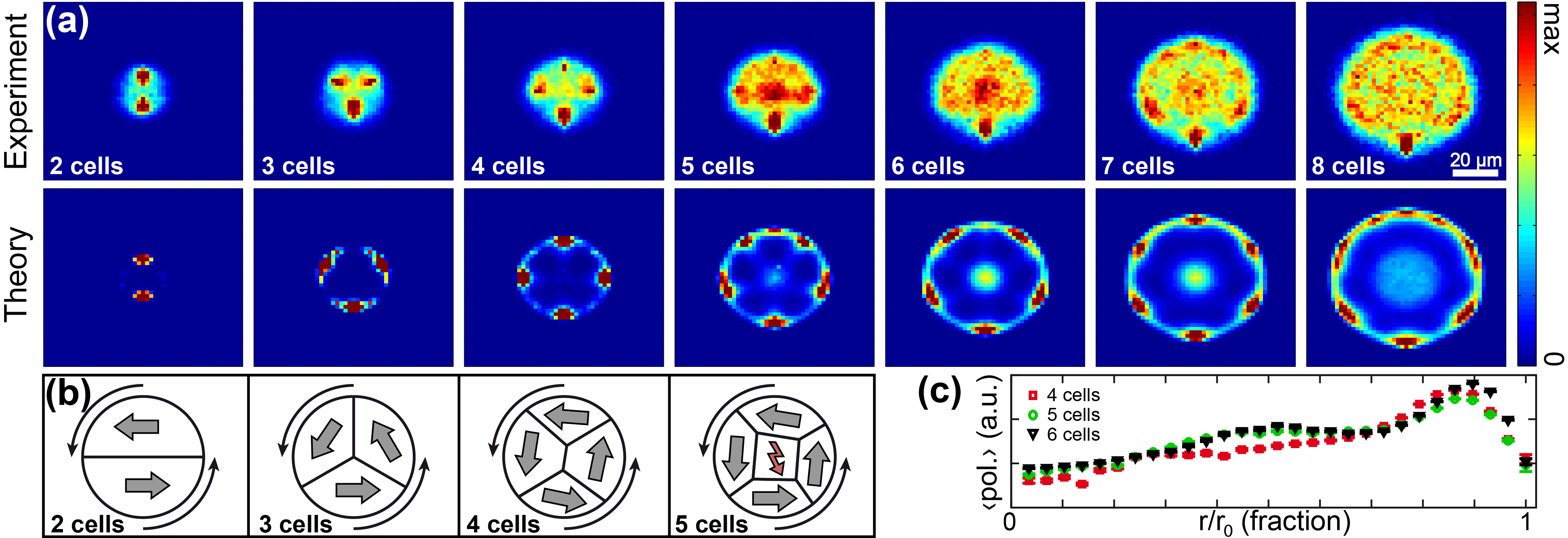}
\caption{\label{Fig.4}(color). (a) Heat map of the relative nuclei positions with respect to a reference nucleus located at the lower border of the system from experimental and theoretical data (for details on the plot generation see section S8 \cite{Supplement}). (b) Schematic of possible polarization alignments during CAMo for different cell numbers. (c) For the CPM, mean magnitude of polarization and its error is plotted against the radial cell position $r$ normalized by the maximal radial cell position $r_0$.
}
\end{figure*}

To further explore the mechanism underlying the discontinuity in persistence time, we monitored the spatial arrangement of cells within the pattern.
Fig.~\ref{Fig.4}(a) shows the relative positions of cells with respect to a reference cell. 
In systems containing up to 4 cells, the cells are predominantly arranged in topologically equivalent positions in the outer regions of the circle. In this configuration, cells in the state of CAMo follow each other in a closed circle. As the number of cells increases to 5, the packing geometry changes abruptly to a conformation in which a single cell is located at the system center.
To connect this topological transition to the observed decrease in the persistence of the CAMo state, intrinsic cell properties have to be accounted for. It is generally assumed that a migrating cell is highly polarized with respect to protein distribution and cytoskeletal organization~\cite{Nishiya:2005, Ridley:2003}. In addition, since neighboring cells are coupled mechanically by cell-cell adhesion, a cell obtains directional guidance cues from adjacent cells. This coupling suggests that adjacent cells tend to align their direction of internal front-rear polarization.
Hence, a ring-like arrangement, as seen for 2-, 3-, and 4-cell systems, naturally provides a stable conformation during a period of CAMo (Fig.~\ref{Fig.4}(b)). If, however, a cell is located in a central position, as in the case of 5 cells, this cell cannot establish a stable axis of internal polarization. 
It seems likely that this lack of orientation leads to the elevated instability we observed for CAMo states of such systems.


To test these heuristic ideas we have developed a computational model~\cite{Thueroff:2014} generalizing the CPM~\cite{Graner:1992, Glazier:1993} to account for both \textit{internal cellular polarization} and \textit{intercellular coupling}.
In the CPM, a cell is represented as a simply connected set of grid sites on a two-dimensional lattice, and thereby cell shape is explicitly represented. The model accounts for mechanical properties of cells and cell-cell adhesion. Previous generalizations of the CPM have implemented cell polarity and ensuing cell migration in a \emph{global} fashion~\cite{Szabo:2010, Kabla:2012} upon adapting ideas from flocking models~\cite{Vicsek:1995}: the overall polarity of a cell is described by a polarity vector, and it is assumed that there is a positive feedback between a cell's displacement and polarity. While these assumptions provide a simple and efficient way to model interactions between a cell and its mechanical environment, they do not resolve internal polarization mechanisms. In fact, there are complex biochemical networks, including Rho family GTPases and membrane lipids, that regulate the assembly of the actin cytoskeleton and thereby the formation of cell protrusions. Recently, computational models have been developed which couple rather sophisticated reaction-diffusion networks to the dynamics of membrane protrusions~\cite{Maree:2006,Maree:2012}. These studies have provided important insights into the spatially resolved signaling processes within cells and how they are affected by cell shape. Here, in order to describe the dynamics off small cell groups, we used an intermediate approach between flocking-type CPM models for cell assemblies~\cite{Szabo:2010,Kabla:2012} and detailed reaction-diffusion models for individual cells~\cite{Maree:2006,Maree:2012}. Specifically, in our computation model we employed an \emph{internal polarization field} within each individual cell to achieve the spatial resolution of microscopic models. At the same time, the numerical algorithm is entirely rule-based (rather than based on complex reaction-diffusion networks) to retain the computational efficiency of CPMs. Furthermore, to account for the effects of cell-cell communication via mechano-transduction, the local dynamics of the internal polarization field is coupled to a cell's membrane protrusions over a finite signaling range. This creates a positive feedback loop integrating intracellular fluctuations and external (mechanical) stimuli and gives rise to spontaneous cell polarization. To match the rotation statistics to the experiments, we simulated cells of fixed (average) size on circular islands at fixed cell density. We then performed a parameter sampling by varying cell adhesion, the range of intracellular mechanical signaling, and the strength of cytoskeletal forces relative to contractile forces. For a more detailed and technical description of the model please refer to S1 in the supplementary material \cite{Supplement}, which also contains a list of the model parameters used.

The model reproduces the symmetry breaking into rotational states found by experiment (see \cite{Supplement} for movies).
Analyzing the numerically generated cell tracks analogously to experimental data, we found CAMo as well as DisMo (see section S7 \cite{Supplement}).
Monte Carlo Steps were adjusted to real time by matching the CAMo peak positions $\bar{\Omega}_N$. We found the same steady decrease of $\bar{\Omega}_N$ with increasing cell number as in the experiments (Fig.~\ref{Fig.3}(c)). Moreover, simulation data also exhibit an increase in CAMo persistence with increasing cell number for 2-,3- and 4-cell systems, and reproduce the discontinuity between 4- and 5-cell systems (Fig.~\ref{Fig.3}(b)). (This feature is also observed when alternative measures for persistence are used; see S6 \cite{Supplement}.)
The discontinuity in persistence is accompanied by the same topological transition in cell arrangement as found by experiment (Fig.~\ref{Fig.4}(a) lower part).
Assessing the characteristics of \emph{internal} cell polarization in the model, we found a systematic decrease of the mean magnitude of the front-rear polarization with decreasing distance to the system center (Fig.~\ref{Fig.4}(c)). 
These findings clearly show that a cell in the center of the pattern is unable to establish a stable axis of polarization and hence destabilizes collective behavior throughout the system. 

Here, we have presented a mesoscopic experimental setup in which the emergence and persistence of collective behavior is analyzed for a small number of cells in confined geometry. 
We showed that it is possible to obtain controlled migrational cell states, which may be classified as disordered and coherent angular motion.
Both experiments and simulations showed consistently that persistence of the coherent state increases with the number of confined cells for small cell numbers but then drops abruptly in a system containing 5 cells. 
This is attributed to a geometric rearrangement of cells to a configuration with a central only weakly polarized cell. It reveals the decisive role of the interplay between local arrangement of neighboring cells and the internal cell polarization in collective migration.
Future studies combining well-controlled cell assemblies confined to micropatterns and computational models may help to evaluate and characterize migrational phenotypes and identify mechanisms that play a key role in cell-to-cell mechano-transduction and finally in the emergence of collective behavior.
\\

\begin{acknowledgments}
This research was supported by the German Excellence Initiative via the program `NanoSystems Initiative Munich' (NIM), and the Deutsche Forschungsgemeinschaft (DFG) via project B01 and project B02 within the SFB 1032.
\end{acknowledgments}

\end{document}